\documentclass[twocolumn]{aastex631}

\graphicspath{{./}{figs/}}
\begin{document}

\title{Effects of Numerical Resolution on Simulated Cloud-Wind Interactions}

\author[0009-0002-0079-4130]{Hannah Leary}
\affiliation{Department of Physics and Astronomy, University of Pittsburgh, 3941 O'Hara St, Pittsburgh, PA 15260}

\author[0000-0001-6325-9317]{Helena M. Richie}
\affiliation{Department of Physics and Astronomy, University of Pittsburgh, 3941 O'Hara St, Pittsburgh, PA 15260}

\author[0000-0001-9735-7484]{Evan Schneider}
\affiliation{Department of Physics and Astronomy, University of Pittsburgh, 3941 O'Hara St, Pittsburgh, PA 15260}

\begin{abstract}

Mixing by hydrodynamical instabilities plays a key role in cloud-wind interactions, causing cloud destruction in the adiabatic limit and facilitating cloud survival with efficient radiative cooling. However, the rate of mixing in numerical simulations is sensitive to the smallest resolved scale, and the relationship between resolution and cloud evolution is under-explored. Using a set of cloud-crushing simulations, we investigate the effects of numerical resolution on cloud survival and acceleration. Modeling both adiabatic and radiative cases, in a subsonic and supersonic wind, we find that cloud survival and velocity does depend on the numerical resolution, however, no single resolution requirement can be applied to all scenarios. In the radiative subsonic case, we find that mass growth and acceleration appear converged at only 4 cells per cloud radius. Conversely, in the supersonic regime, we see a clear dependence of cloud destruction and velocity on resolution that is not converged even at 48 cells per cloud radius, implying that accurately capturing cloud destruction may require higher resolution than capturing growth. We also present a simple model illustrating how ram pressure accelerates cool clouds at early times before mixing kicks in as an acceleration mechanism.

\end{abstract}


\section{Introduction} \label{sec:intro}

Galactic outflows are a crucial component of the baryon cycle that drives galaxy formation and evolution. Outflows regulate the mass and star formation rate of a galaxy by sweeping out gas from potential star forming regions in the interstellar medium (ISM) \citep{Dekel1986, Scannapieco2008, Vogelsberger2014, Schaye2015, Dave2016}, and can carry this gas into the circumgalactic medium (CGM) and intergalactic medium (IGM), thereby enriching them with metals \citep{Scannapieco2002, Oppenheimer2006, Ford2014, Hafen2019}. Stellar feedback is thought to be a major driver of galactic outflows, including feedback from star formation and supernova explosions \citep{Strickland2009, Hopkins2012, Somerville2015}. The precise impact that the baryon cycle has on galaxy evolution is not well understood, so it is crucial to study the processes by which gas is driven out of galaxies.

 Because of their importance in galaxy evolution, accurately modeling these outflows in numerical simulations is a major focus of current studies. Most cosmological simulations rely on adaptive mesh refinement or Lagrangian methods to model processes of varying scales in galaxies. While this is effective at reducing computational cost and resolving dense gas, these simulations often under-resolve hot winds and face challenges in resolving the hydrodynamic processes and instabilities that occur during the interaction between different temperature phases in outflows. Thus, one goal of recent theoretical work is to develop more accurate models of phase interactions in galactic winds that may be used as sub-grid prescriptions in cosmological simulations \citep{Kim2020, Huang2022, Fielding2022, Butsky2024, Smith2024}. 

The gas within outflows is observed to be multiphase, with molecular, cold, ionized, and hot phases detected across the electromagnetic spectrum \citep[see reviews by][]{Veilleux2005, Rupke2018}. The energy budget of these winds is dominated by hot ($\geq 10^6~\textrm{K}$) diffuse gas, while the mass budget is dominated by cool ($\leq 10^4~\textrm{K}$), denser gas. This range can be even more extreme in very high pressure environments like the base of winds or massive systems, with hot phase temperatures reaching $\geq10^7$ K, thus leading to higher overdensities of the cool phase \citep{Fabian2012}. Blueshifted line-of-sight absorption studies show that this cool gas is moving out of galaxies at velocities ranging up to 1000 km/s \citep{Rupke2005, Martin2006, Weiner2009, Rubin2014}, i.e. it is likely ``entrained" in the hot wind. The focus of many numerical studies has been to understand how cool gas clouds can be accelerated to hundreds of km/s without being heated and mixed into the hot phase, or ``destroyed".

The classic picture of how cool clouds reach observed velocities is one where the hot wind ram pressure accelerates the clouds. For a cloud at rest in a wind of density $\rho_\mathrm{w}$ and velocity $v_\mathrm{w}$, the ram pressure of the wind is

\begin{equation}
P_{\rm ram} \sim \rho_\mathrm{wind}v^2_\mathrm{wind}.
\label{eq:P_ram}
\end{equation}

\noindent Thus, the force acting on a cloud of radius $R_\mathrm{cl}$ can be approximated by

\begin{equation}
F_\mathrm{ram} \sim \pi R_\mathrm{cl}^2P_\mathrm{ram}.
\label{eq:F_ram}
\end{equation}

\noindent The acceleration due to ram pressure is then given by

\begin{equation}
\frac{dv_\mathrm{cl}}{dt} \sim \frac{\rho_\mathrm{w} v_\mathrm{w}^2}{\rho_\mathrm{cl} R_\mathrm{cl}},
\label{eq:ramaccel}
\end{equation}

\noindent so 
\begin{equation}
t_{\rm{ent, ram}} = \frac{\rho_{\rm{cl}} R_{\rm{cl}}}{\rho_{\rm{w}} v_{\rm{w}}}
\label{eq:t_ent}
\end{equation}

\noindent is a rough estimate of the time it would take to accelerate clouds to the wind velocity. While ram pressure acceleration likely plays a significant role in the early stage of acceleration, high resolution cloud-wind simulations have shown that ram pressure alone is not sufficient for cloud entrainment, as hydrodynamic instabilities such as Kelvin-Helmholtz (KH) destroy clouds in several characteristic ``cloud crushing times" \citep{Klein1994, Xu1995, Zhang2017}. For a cloud-wind density contrast $\rho_{\rm cl}/\rho_{\rm w} = \chi$, this timescale is given by

\begin{equation}
t_\mathrm{cc}=\chi^{1/2} \frac{ R_\mathrm{cl}}{v_\mathrm{w}}.
\label{eq:tcc_adiabatic}
\end{equation}
\noindent Thus, $t_\mathrm{cc} < t_\mathrm{ent, ram}$, hence, the ``entrainment problem."

More recent works have shown that the mixing induced by instabilities can actually lead to cloud survival and growth when the mixed gas is able to cool radiatively on a timescale that is short compared to the cloud crushing time , i.e., when $t_{\rm cool,mix} < t_{\rm cc}$, where $t_{\rm cool,mix}$ is given by

\begin{equation}
t_{\rm cool,mix} = \frac{k_{\rm B}T_{\rm mix}}{n_{\rm mix}\Lambda(T_{\rm mix})}.
\label{eq:tcool}
\end{equation}

\noindent and $k_{\rm B}$ is the Boltzmann constant. The gas temperature $T_{\rm mix}$ and number density $n_{\rm mix}$ in these turbulent radiative mixing layers (TRMLs) can be roughly characterized by the geometric mean of the cold and hot phases' temperatures and number densities. The cooling rate $\Lambda$ roughly peaks around the temperature of the mixed gas ($\sim 10^5 \ \rm K$),making the mixed gas highly efficient at cooling. The mixed material thus can rapidly cool and accrete onto the cloud, enabling long-term survival and acceleration through the transfer of mass and momentum. This model predicts cloud entrainment in the fast-cooling regime, where $t_{\rm cool,mix} \ll t_{\rm cc}$, potentially explaining the high-velocity cool gas often observed in winds \citep{Gronke2018, Abruzzo2022, Fielding2022}. In particular, this model suggests that the velocity of the cool material in winds is primarily achieved through momentum transfer via mixing, leading to a linear relationship between the fraction of condensed hot wind material and the velocity of cool gas \citep{Schneider2020}.

Mixing thus plays a crucial role in cold cloud evolution, causing destruction in the adiabatic limit where gas cannot lose energy through radiative cooling, and facilitating survival in the radiative case. However, because the growth rate of KH instabilities at a discontinuous interface in simulations is sensitive to the smallest resolved scale \citep[e.g.][]{Chandrasekhar1961, Robertson2010}, the rate of mixing may increase as the resolution of the simulation is increased. Indeed, previous work has demonstrated that as simulation resolution increases, finer structures appear and the number of cloudlets increases, thus substantially increasing the cloud surface area available for mixing \citep{Cooper2009, Schneider2017}. 

Despite mixing playing such an important role in cloud evolution, the relationship between cloud evolution and numerical resolution is an under-explored area, particularly in the low resolution limit most relevant in cosmological simulations. This study aims to quantify the relationship between numerical resolution and cloud acceleration and survival. Clarifying resolution effects will help confirm if cosmological simulations can accurately capture this mixing and cooling model, or if subgrid models for cloud-wind interactions are necessary. Additionally, determining the extent of the model's dependence on numerical resolution will help to confirm that it is sufficient to explain observed properties of entrained clouds. This is necessary for understanding if additional acceleration mechanisms, such as cosmic rays or radiation pressure, are required to explain cloud evolution in winds. 

In this work, we explore the effects of numerical resolution on cloud evolution in both the adiabatic and radiative cases using a spherical cloud in an idealized wind tunnel setup. We also investigate the role of wind speed and ram pressure acceleration by modeling both the supersonic and the less commonly studied subsonic wind regime. In Section \ref{sec:methods} we provide the parameters of our simulations and details of the analysis, and we present the results in Section \ref{sec:results}. In Section \ref{sec:disc} we discuss the relevance of ram pressure acceleration at early times as well as the implications of resolution effects in the context of modern cosmological simulation suites.  

\begin{deluxetable*}{ccccccccc}[ht!]
\tablenum{1}
\label{tab:sims}
\tablewidth{0pt}
\tablehead{
\colhead{Group} & \colhead{Resolution} & \colhead{Dimensions ($R_{\rm cl}$)} & \colhead{$v_{\rm w}$~(km/s)} & \colhead{$t_{\rm cc}$ (kyr)} & \colhead{$t_{\rm cool,mix}$ (kyr)} & \colhead{$t$ ($t_{\rm cc}$)} & \colhead{M} & \colhead{Model}\\
}
\startdata
a100 & $R_\mathrm{cl}$/4 & $32\times16\times16$ & 100 & $4.89\times10^{3}$ & n/a & 10 & 0.659 & Adiabatic\\
& $R_\mathrm{cl}$/8  & \texttt{"} & \texttt{"}  & \texttt{"} & \texttt{"} & \texttt{"} & \texttt{"} & \texttt{"}\\
& $R_\mathrm{cl}$/16  & \texttt{"} & \texttt{"}  & \texttt{"} & \texttt{"} & \texttt{"} & \texttt{"} & \texttt{"}\\
& $R_\mathrm{cl}$/32  & \texttt{"} & \texttt{"}  & \texttt{"} & \texttt{"} & \texttt{"} & \texttt{"} & \texttt{"}\\
& $R_\mathrm{cl}$/48  & \texttt{"} & \texttt{"}  & \texttt{"} & \texttt{"} & \texttt{"} & \texttt{"} & \texttt{"}\\
\hline
a1000 & $R_\mathrm{cl}$/4 & $32\times16\times16$ & 1000 & $4.89\times10^{2}$ & n/a & 10 & 6.59 & Adiabatic\\
& $R_\mathrm{cl}$/8 & \texttt{"} & \texttt{"}  & \texttt{"} & \texttt{"} & \texttt{"} & \texttt{"} & \texttt{"}\\
& $R_\mathrm{cl}$/16 & \texttt{"} & \texttt{"}  & \texttt{"} & \texttt{"} & \texttt{"} & \texttt{"} & \texttt{"}\\
& $R_\mathrm{cl}$/32 & \texttt{"} & \texttt{"}  & \texttt{"} & \texttt{"} & \texttt{"} & \texttt{"} & \texttt{"}\\
& $R_\mathrm{cl}$/48 & \texttt{"} & \texttt{"}  & \texttt{"} & \texttt{"} & \texttt{"} & \texttt{"} & \texttt{"}\\
\hline
r100 & $R_\mathrm{cl}$/4 & $48\times16\times16$ & 100 & $4.89\times10^{3}$ & 9.39 & 10 & 0.659 & Radiative Cooling\\
& $R_\mathrm{cl}$/8 & \texttt{"} & \texttt{"}  & \texttt{"} & \texttt{"} & \texttt{"} & \texttt{"} & \texttt{"}\\
& $R_\mathrm{cl}$/16 & \texttt{"} & \texttt{"}  & \texttt{"} & \texttt{"} & \texttt{"} & \texttt{"} & \texttt{"}\\
& $R_\mathrm{cl}$/32 & \texttt{"} & \texttt{"}  & \texttt{"} & \texttt{"} & \texttt{"} & \texttt{"} & \texttt{"}\\
& $R_\mathrm{cl}$/48 & \texttt{"} & \texttt{"}  & \texttt{"} & \texttt{"} & \texttt{"} & \texttt{"} & \texttt{"}\\
\hline
r1000 & $R_\mathrm{cl}$/4 & $48\times16\times16$ & 1000 & $4.89\times10^{2}$ & 9.39 & 20 & 6.59 & Radiative Cooling\\
& $R_\mathrm{cl}$/8 & \texttt{"} & \texttt{"}  & \texttt{"} & \texttt{"} & \texttt{"} & \texttt{"} & \texttt{"}\\
& $R_\mathrm{cl}$/16 & \texttt{"} & \texttt{"}  & \texttt{"} & \texttt{"} & \texttt{"} & \texttt{"} & \texttt{"}\\
& $R_\mathrm{cl}$/32 & \texttt{"} & \texttt{"}  & \texttt{"} & \texttt{"} & \texttt{"} & \texttt{"} & \texttt{"}\\
& $R_\mathrm{cl}$/48 & \texttt{"} & \texttt{"}  & \texttt{"} & \texttt{"} & \texttt{"} & \texttt{"} & \texttt{"}\\
\hline
\enddata
\caption{Parameters for simulations discussed in Section \ref{sec:methods}. Resolution refers to the number of cells per cloud radius, dimensions is the x, y, and z lengths of the simulation box (in units of initial cloud radius), $v_{\rm w}$ is the velocity of the wind in km/s, $t_{\rm cc}$ is the cloud-crushing time in kyr, $t_{\rm cool,mix}$ is the cooling time of the mixed gas in kyr, $t$ is the total simulation time in cloud-crushing times, and M is the Mach number of the wind (M = $v_{\rm w} / c_{\rm s,w}$)}
\end{deluxetable*}

\section{Simulations and Analysis} \label{sec:methods}

To investigate the effects of resolution on cloud evolution, we ran four sets of simulations at five different resolutions: $R_{\rm cl}/\Delta x$ = 4, 8, 16, 32, and 48 (hereafter $R_{4}$, $R_{8}$, $R_{16}$, $R_{32}$, $R_{48}$). Table \ref{tab:sims} shows the parameters for our sets of simulations: a100, a1000, r100, and r1000. Adiabatic simulations begin with an ``a" and simulations with radiative cooling begin with an ``r", and the following number denotes the wind velocity in $\mathrm{km/s}$. All simulations were run with the Cholla hydrodynamics code \citep{Schneider2015} using PPMP reconstruction \citep{Colella1984}, the HLLC Riemann solver \citep{Toro1994, Batten1997}, and the Van Leer integrator \citep{Stone2009}. Our radiative simulations employ a piecewise parabolic fit to a collisional ionization equilibrium cooling curve for solar metallicity gas, with a cutoff below $10^4$~K \citep{Ferland2013, Schneider2018}. We employ a standard wind tunnel setup consisting of a long box with a constant wind entering from the left boundary and outflow boundaries on all other sides. The adiabatic simulations have box dimensions of $32~R_{\rm cl}\times16~R_{\rm cl}\times16~R_{\rm cl}$, while the radiative simulations have extended box dimensions of $48~R_{\rm cl}\times16~R_{\rm cl}\times16~R_{\rm cl}$, to account for the fact that cloud material survives for longer with radiative cooling.

In order to explore potential differences in mixing for the subsonic versus supersonic regime, we use wind speeds of 100 km/s (M = 0.659) for a100 and r100, and 1000 km/s (M = 6.59) for a1000 and r1000, respectively. The wind density and temperature are the same for all simulations: $n_{\rm w} = 0.01$ cm$^{-3}$ and $T_{\rm w} = 10^{6}$ K. Clouds are initialized as spheres with a radius of 50 pc and zero initial velocity, and are positioned 3.2 $R_{\rm cl}$ away from the $-x$ boundary. The clouds are initially in thermal pressure equilibrium with the wind, with a cloud-wind density contrast of $\chi$ = $10^2$, and a cloud temperature of $T_{\rm cl} = 10^4$~K. All simulations were run for a total of 10 $t_{\rm cc}$ except for r1000, which runs for 20 $t_{\rm cc}$. The cloud-crushing times (from Equation \ref{eq:tcc_adiabatic}) are $4.89\times 10^3$ kyr and $4.89\times 10^2$ kyr for the subsonic and supersonic simulations, respectively. The cooling time of the mixed gas (from Equation \ref{eq:tcool}) in the radiative simulations is 9.39 kyr, giving a ratio $t_{\rm cool,mix}/t_{\rm cc}$ of $1.92\times 10^{-3}$ for a100 and $1.92\times 10^{-2}$ for a1000. All simulation parameters are listed in Table \ref{tab:sims}.

\begin{figure*}[t]
\begin{interactive}{animation}{movies/a100.mp4}
\centering
\includegraphics[width=.66\linewidth]{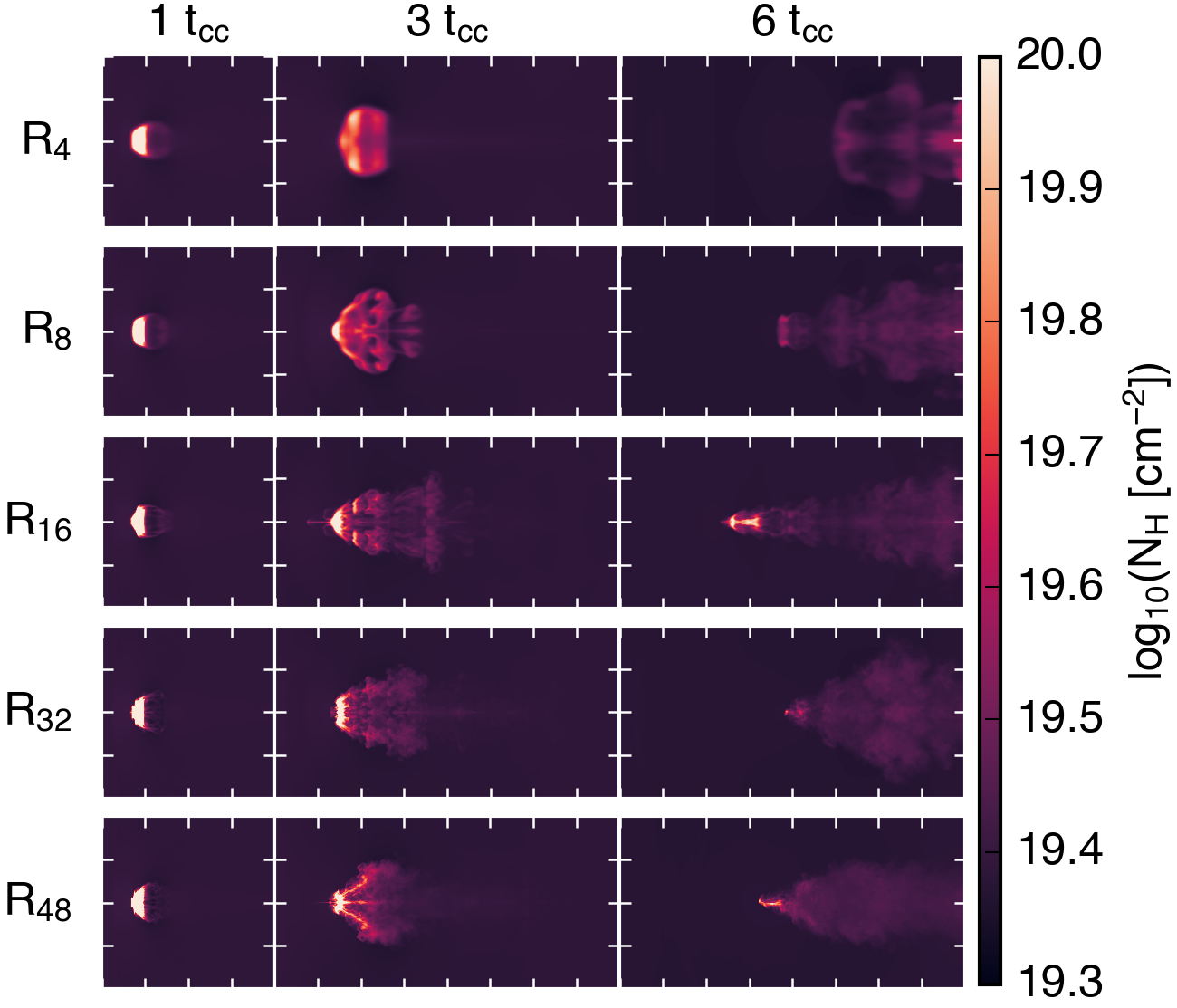}
\end{interactive}
\begin{interactive}{animation}{figs/a1000.mp4}
\includegraphics[width=.66\linewidth]{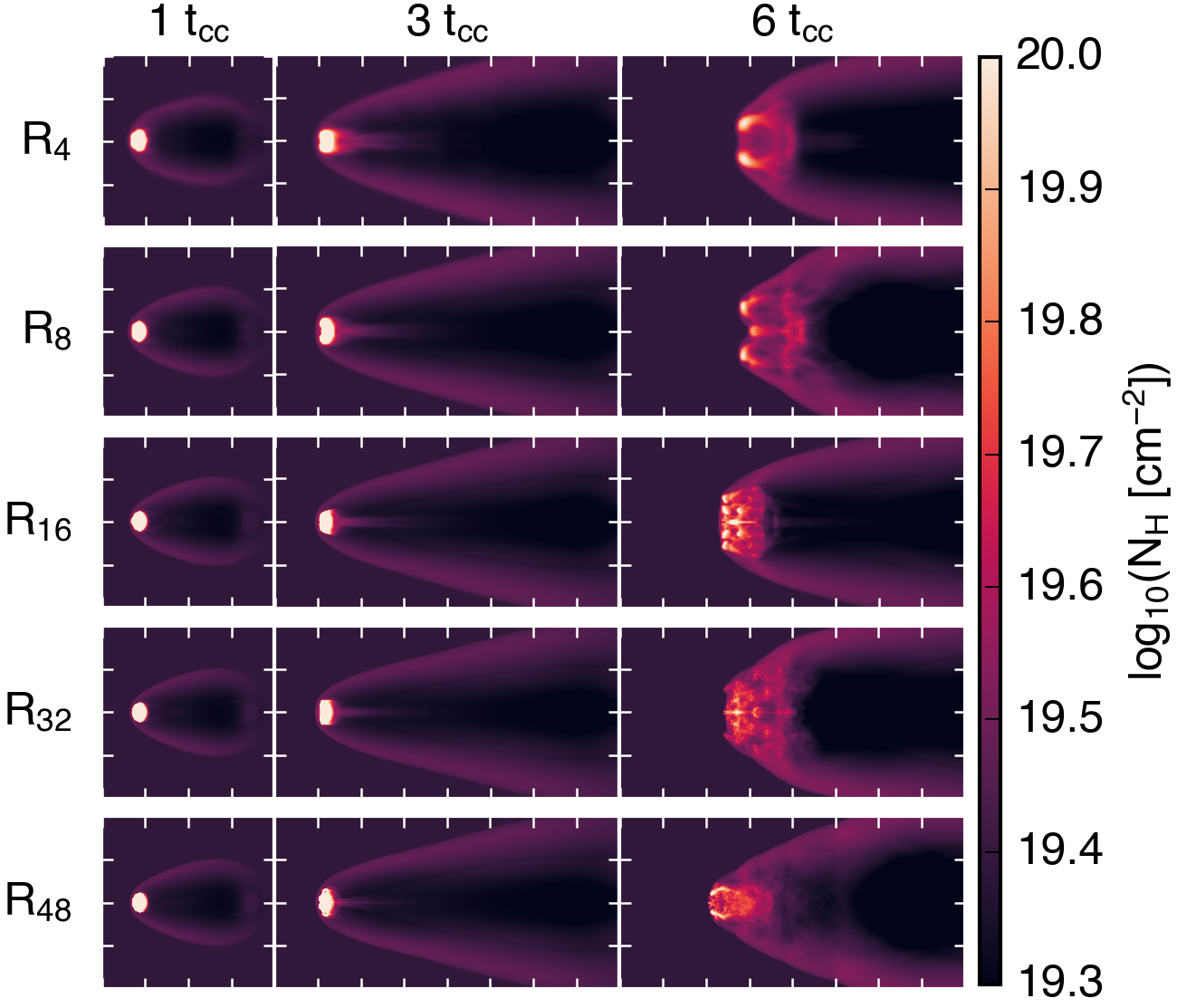}
\end{interactive}
\caption{Snapshots of column density for adiabatic cloud-wind simulations with resolutions $R_{\rm cl}/\Delta x$ = 4, 8, 16, 32, and 48. Top and bottom panels depict the evolution of a cloud in the subsonic (a100) and supersonic (a1000) winds, respectively. In the a100 simulation, all clouds are quickly disrupted by the wind as they move along the box, with the exception of $R_{\rm 16}$, which remains far behind the other resolutions throughout the simulation. In the a1000 simulation, a bowshock is visible as the wind interacts with the cloud. In all resolutions, the clouds are similarly disrupted as they are accelerated along the box.} \label{fig:adiabatic snapshots}
\end{figure*}

In the following analysis, we consider cells with a density above $1/3$ of the initial cloud density to be ``cloud" material\footnote{Our results are not sensitive to the precise density threshold; see Appendix \ref{sec:cuts}}. We measure cloud mass,
\begin{equation}
M_{\rm cl} = \sum \rho_i(> \frac{1}{3}\rho_\mathrm{cl,init}) \mathrm{dV},
\label{eq:M_cl}
\end{equation}
and mass-weighted average cloud velocity in the wind direction,
\begin{equation}
\bar{v}_x=\frac{ \sum\rho_i(> \frac{1}{3}\rho_\mathrm{cl,init}) v_{x, i}\mathrm{dV}}{M_{\rm cl}},
\label{eq:v_avg}
\end{equation}
every 0.2 $t_{\rm cc}$ (0.4 $t_{\rm cc}$ for r1000), where $\rho_i$ is the mass density of the $i$th cell and $\mathrm{dV}$ is the cell volume.

\section{Results} \label{sec:results}

\begin{figure}[t]
    \centering
    \includegraphics[width=\linewidth]{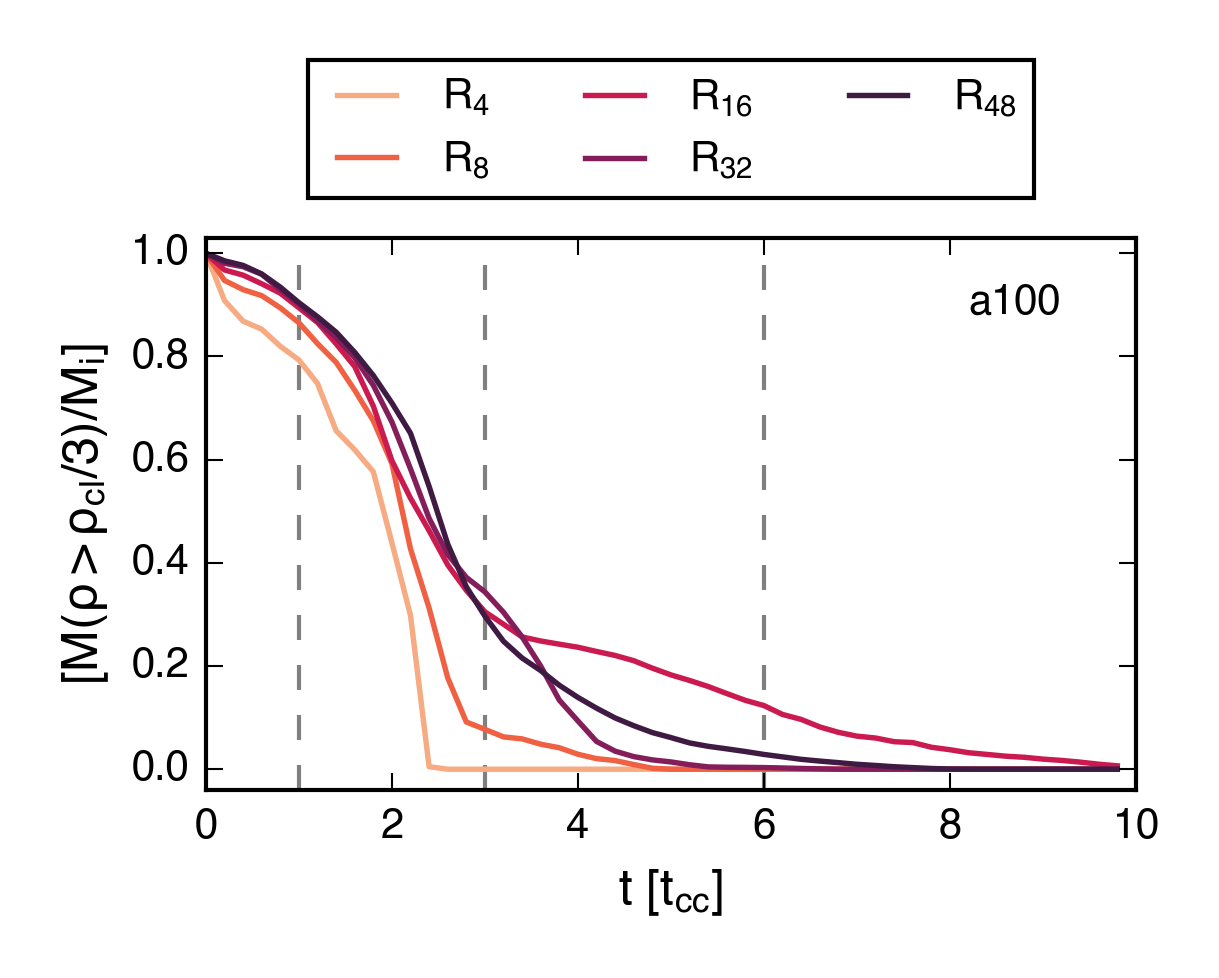}
    \includegraphics[width=\linewidth]{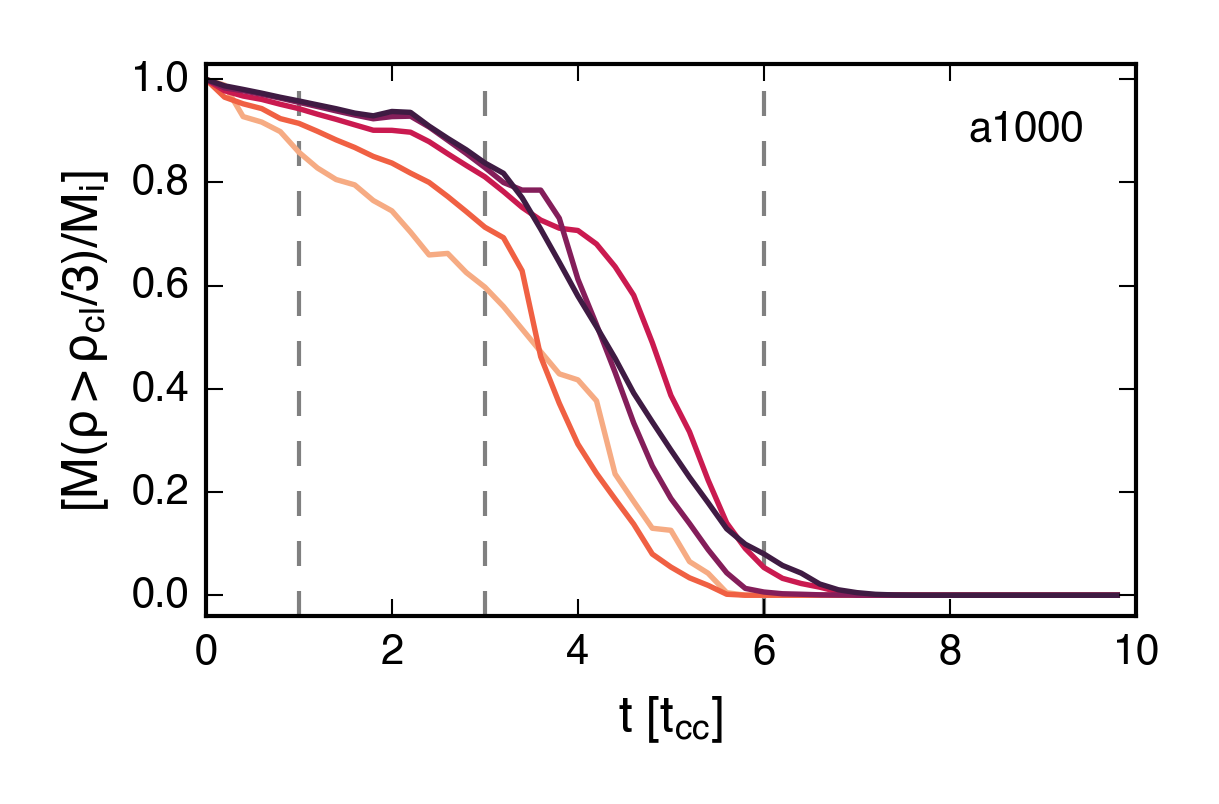}
    \caption{Change in cloud mass over time normalized by the initial cloud mass for resolutions $R_{\rm cl}/\Delta x$ = 4, 8, 16, 32, and 48. Top and bottom panels depict adiabatic subsonic and supersonic models a100 and a1000, respectively. Vertical dashed lines correspond to snapshots shown in Figure \ref{fig:adiabatic snapshots}.}\label{fig:adiabatic mass}
\end{figure}

\begin{figure}[t]
    \centering
    \includegraphics[width=\linewidth]{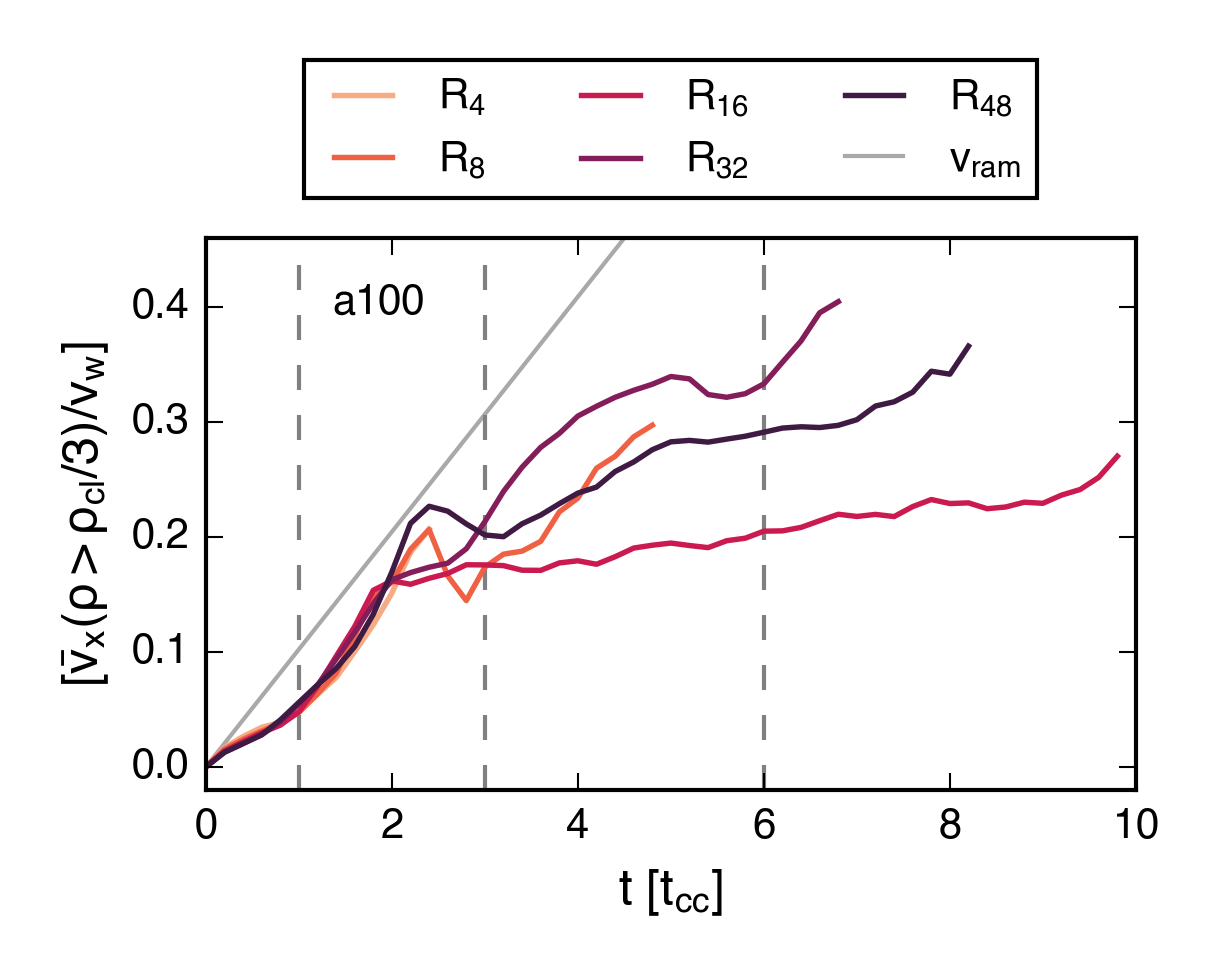}
    \includegraphics[width=\linewidth]{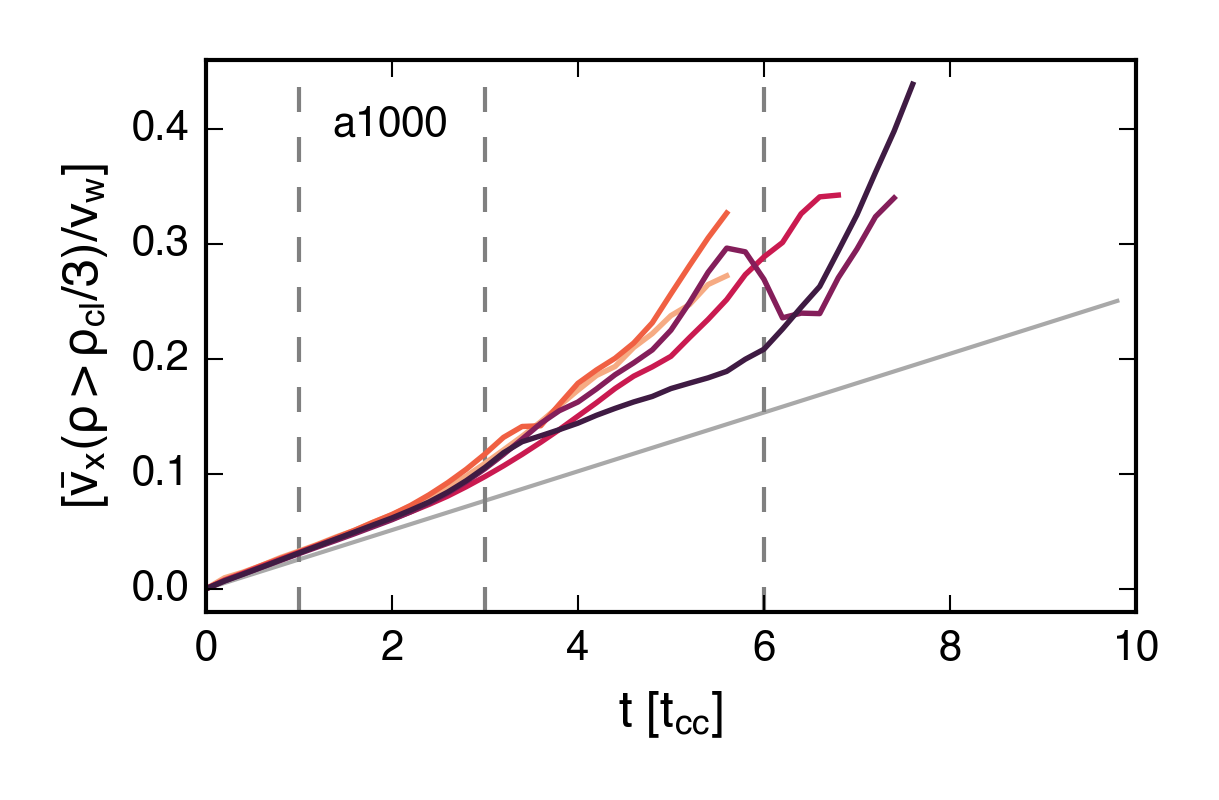}
    \caption{Change in cloud velocity over time normalized to the wind velocity for resolutions $R_{\rm cl}/\Delta x$ = 4, 8, 16, 32, and 48. Top and bottom panels depict models a100 and a1000, respectively. Vertical dashed lines correspond to snapshots shown in Figure \ref{fig:adiabatic snapshots}. The solid gray line in each panel indicates the expected cloud velocity due to ram pressure acceleration.} \label{fig:adiabatic velocity}
\end{figure}

In each of the four cloud-wind models, the resolution trends vary significantly. In the adiabatic simulations, the trends are surprisingly non-monotonic, with the lowest and highest resolutions behaving most similarly. On the other hand, the radiative simulations follow a more linear pattern and in the subsonic case, cloud evolution appears consistent across all resolutions. 

\subsection{Adiabatic Subsonic Models}

Figure \ref{fig:adiabatic snapshots} illustrates the non-monotonic trend in the adiabatic case. The panels show column density snapshots of the cloud at $1 ~t_{\rm cc}$, in a partially disrupted state at 3 $t_{\rm cc}$, and late in its evolution at 6 $t_{\rm cc}$. These snapshots illustrate how hydrodynamic instabilities mostly destroy the cloud within $\sim 6~t_{\rm cc}$, as expected. In both the subsonic and supersonic regimes, the cloud's acceleration and destruction are affected by the resolution of the simulation. In the a100 simulations in particular (top panel), the destruction follows an almost parabolic trend, with $R_{16}$ lagging behind the most, and the lowest and highest resolutions accelerating more quickly. This behavior is clearest at $6~t_{\rm cc}$, when all clouds are nearly destroyed except for $R_{16}$, which still retains a significant amount of cloud material and is much farther behind the other clouds. 

The behavior of a100 can be seen more quantitatively in the top panel of Figure \ref{fig:adiabatic mass}, which shows the normalized cloud mass, $M_{\rm cl}/M_{\rm cl,i}$ over time. In the first few $t_{\rm cc}$, the remaining cloud mass monotonically decreases with decreasing resolution, with the highest resolutions retaining the most mass, and the lowest resolutions losing mass the fastest; by $2~t_{\rm cc}$, the $R_{4}$ cloud has lost over 50\% of its initial mass, while the $R_{48}$ cloud still retains over 70\%. At $\sim2~t_{\rm cc}$, however, the resolution trend breaks down, with $R_{16}$ exceeding the higher resolutions in mass, and this model continues to retain material above the density threshold until $10~t_{\rm cc}$. All other resolutions are completely destroyed by $7~t_{\rm cc}$, and the lowest resolutions are disrupted notably faster than the others---$R_{4}$ is completely destroyed before $3~t_{\rm cc}$. 

This mass evolution is consistent with the average velocities of cloud material, as shown in the top panel of Figure \ref{fig:adiabatic velocity}, where we plot the average cloud velocity in the wind direction normalized by the wind velocity, $\bar{v}_{x}/v_{\rm w}$, as a function of time. At early times, the models all accelerate extremely similarly. However, after $2~t_{\rm cc}$, we see that the cloud material in $R_{16}$ accelerates more slowly than the other models, and reaches a fairly consistent terminal velocity of only $\sim20\%$ of the wind speed before being completely destroyed. In contrast, the surviving cloud material in other models accelerates to $30 - 40\%$ of the wind velocity, and never reaches a terminal velocity before being completely destroyed.

As noted above, the velocities of all five resolutions are nearly identical until $\sim 2~t_{\rm cc}$, when they begin to drastically diverge from one another. This can be explained by the different acceleration mechanisms at play at different stages in the cloud evolution. In the first few $t_{\rm cc}$ before hydrodynamic instabilities have grown enough to disrupt the cloud, ram pressure dominates the acceleration. We demonstrate this by plotting the expected velocity from ram pressure acceleration, calculated from Equation \ref{eq:ramaccel}. The initial cloud velocities are only slightly lower than the expected value, and the overall slope of the lines are quite similar. The slight discrepancy is likely due to our criterion for cloud material ($\rho \geq \rho_{\rm cl,i}/3$); some momentum is transferred to cloud material which is too diffuse to make the cut, effectively lowering the overall average velocities (see Appendix \ref{sec:cuts}). At later times, mixing begins to affect the acceleration, as demonstrated by the different velocity trends for different resolution models. All of the models have velocities that are significantly below the ram pressure expectation, but this behavior can again be explained by our density cutoff -- mixed gas at the cloud-wind interface that has a higher fraction of wind material will have higher velocities, but will also have lower density, so using a density cutoff to define cloud material biases our analysis towards tracking the lower-velocity, denser gas.

\subsection{Adiabatic Supersonic Models}

The second panel of Figure \ref{fig:adiabatic snapshots} shows snapshots of the supersonic adiabatic simulations. Here the cloud evolution is qualitatively different from the subsonic case. The interaction of the supersonic flow with the initial cloud results in a bow shock, which is characterized by a discontinuous change in density, velocity, and pressure. Thus, the cloud experiences these post-shock conditions instead of the initialized wind conditions that the cloud in the subsonic wind feels. Additionally, high Mach numbers are known to suppress the KH instability \citep{Madelker2016}.

Indeed, we see that the resolution dependence here is subtler than in the subsonic case, with all five resolutions appearing to have a more similar evolution by $6~t_{\rm cc}$. However, the bottom panel of Figure \ref{fig:adiabatic mass} shows that the early time mass evolution is similar to the subsonic case. Mass increases monotonically with increasing resolution until $4~t_{\rm cc}$, when $R_{16}$ takes the lead in mass. The lowest resolution models are again destroyed the fastest, with over 50\% of cloud material lost by $4~t_{\rm cc}$. In contrast with the subsonic regime, however, all resolutions reach total destruction within $2~t_{\rm cc}$ of each other, and all are destroyed by $7~t_{\rm cc}$, despite taking longer to lose mass at early times. In Figure \ref{fig:adiabatic velocity} (bottom), we again see very similar acceleration between all of the models until $3~t_{\rm cc}$, after which the velocities begin to diverge, and the lowest resolution simulations appear to accelerate the fastest. In this regime, none of the models reach a consistent terminal velocity before being destroyed, and the highest velocities achieved are again 30 - 40\% of the wind velocity. In both the subsonic and supersonic regimes, the highest speeds are achieved in the high resolution simulations, indicating that increased mixing as a result of higher numerical resolution is likely at play.

As in the subsonic case, we plot in Figure \ref{fig:adiabatic velocity} a line representing the expected velocity due to ram pressure acceleration. However, in this supersonic case we consider the adiabatic post-shock density and velocity for the wind. For strong shocks, the post-shock density increases by a factor of 4 and the velocity decreases by the same factor. Equation \ref{eq:ramaccel} thus yields a decrease in the ram pressure force by a factor of 4, so our expected ram pressure velocity is decreased by a factor of 4 compared to the subsonic regime. Indeed, the clouds closely follow this velocity for the first $t_{\rm cc}$ before accelerating at a faster rate, again indicating that acceleration due to mixing begins to play a role at later times.

\subsection{Radiative Subsonic Models}

\begin{figure*}
\begin{interactive}{animation}{r100.mp4}
\centering
\includegraphics[width=0.78\linewidth]{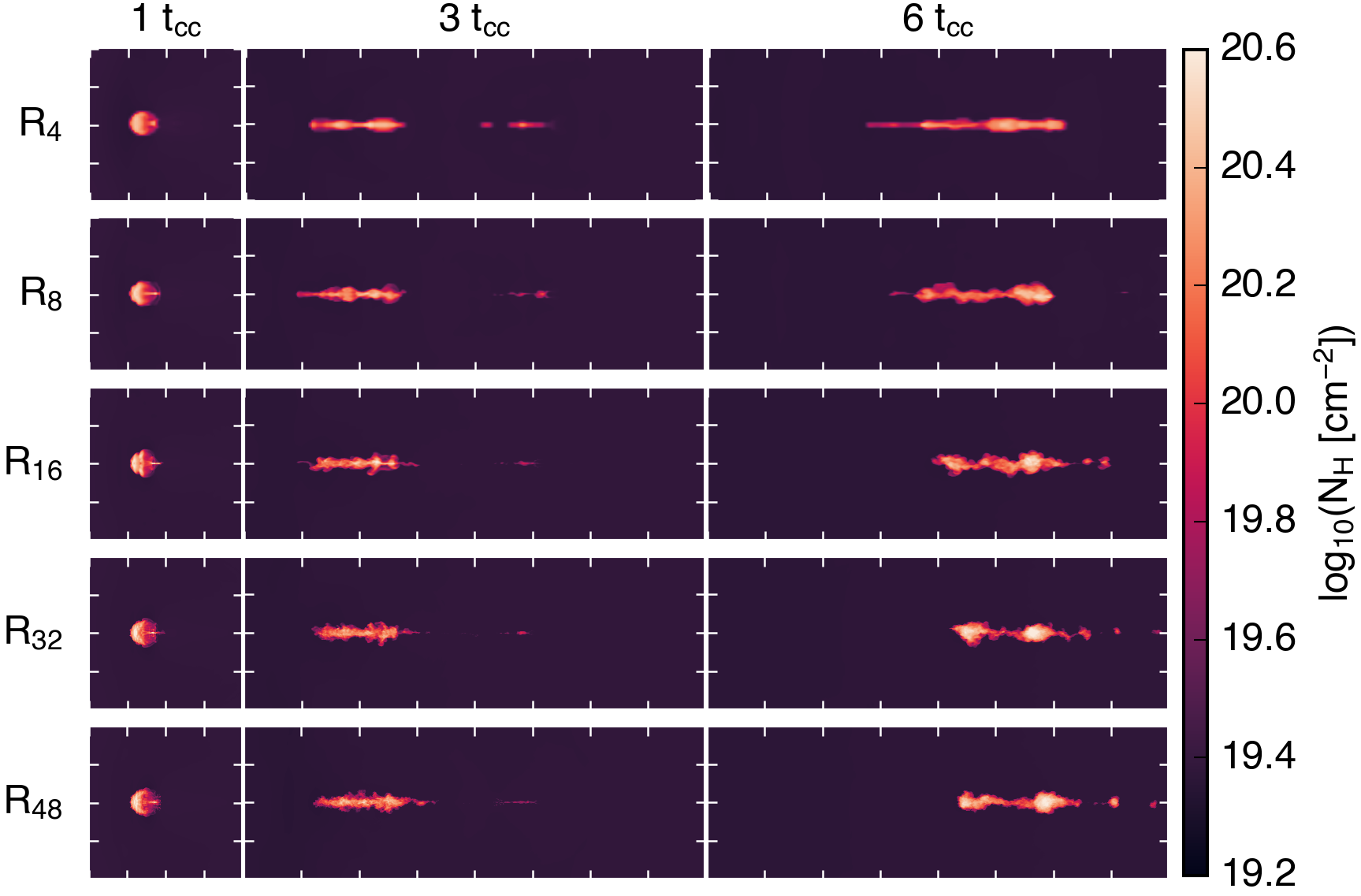}
\end{interactive}
\begin{interactive}{animation}{r1000.mp4}
\includegraphics[width=0.78\linewidth]{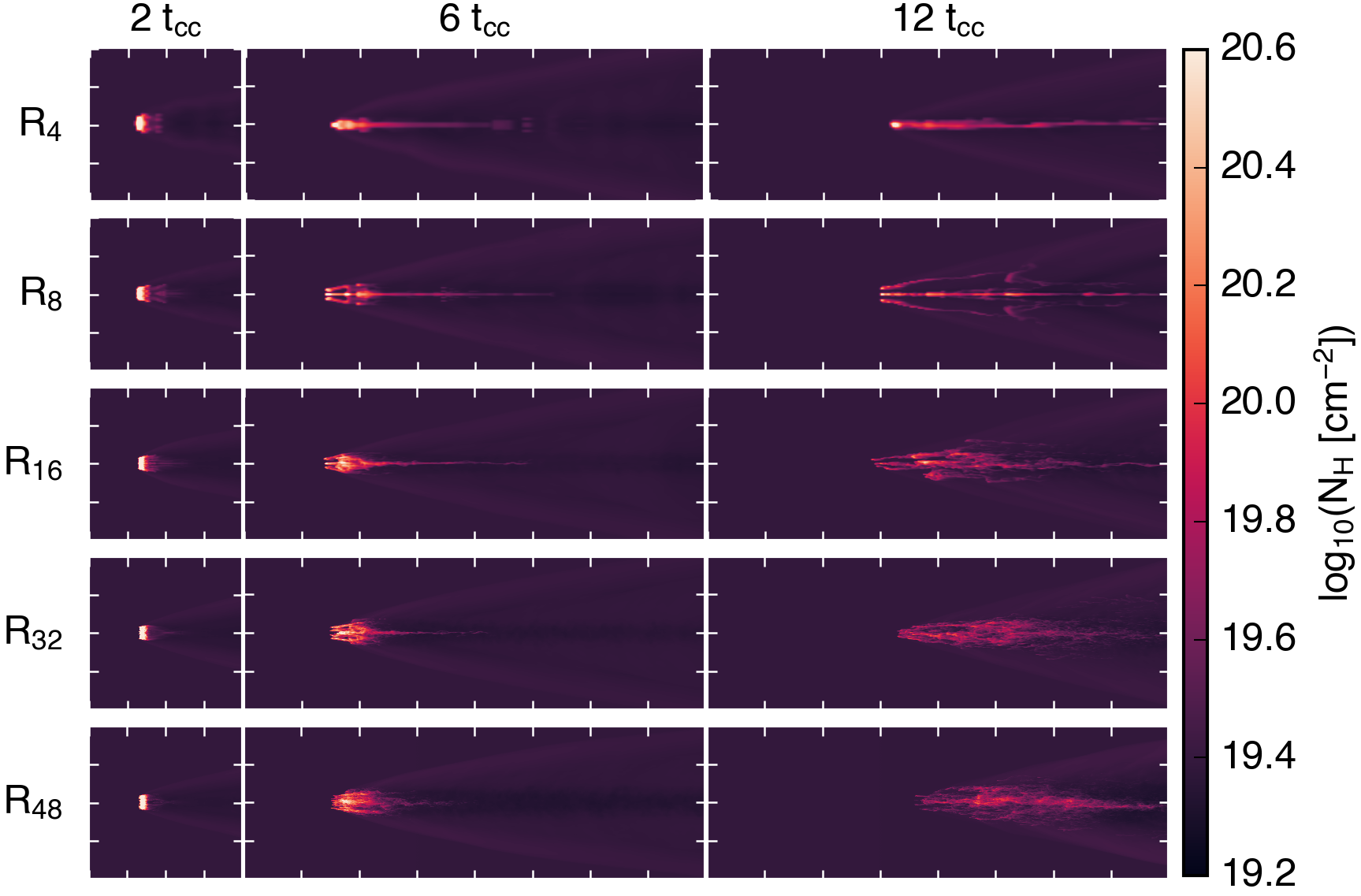} 
\end{interactive}
\caption{Snapshots of column density at resolutions $R_{\rm cl}/\Delta x$ = 4, 8, 16, 32, and 48. Top and bottom panels show the evolution of radiative clouds in the r100 and r1000 simulations, respectively. The clouds in the r100 model traverse the box for 10 $t_{\rm cc}$ and retain a clumpy, narrow structure. The r1000 clouds move along the box for a total of 20 $t_{\rm cc}$ and undergo varying degrees of fragmentation. }

\label{fig:radiative snapshots}
\end{figure*}

\begin{figure}[t!]
    \includegraphics[width=\linewidth]{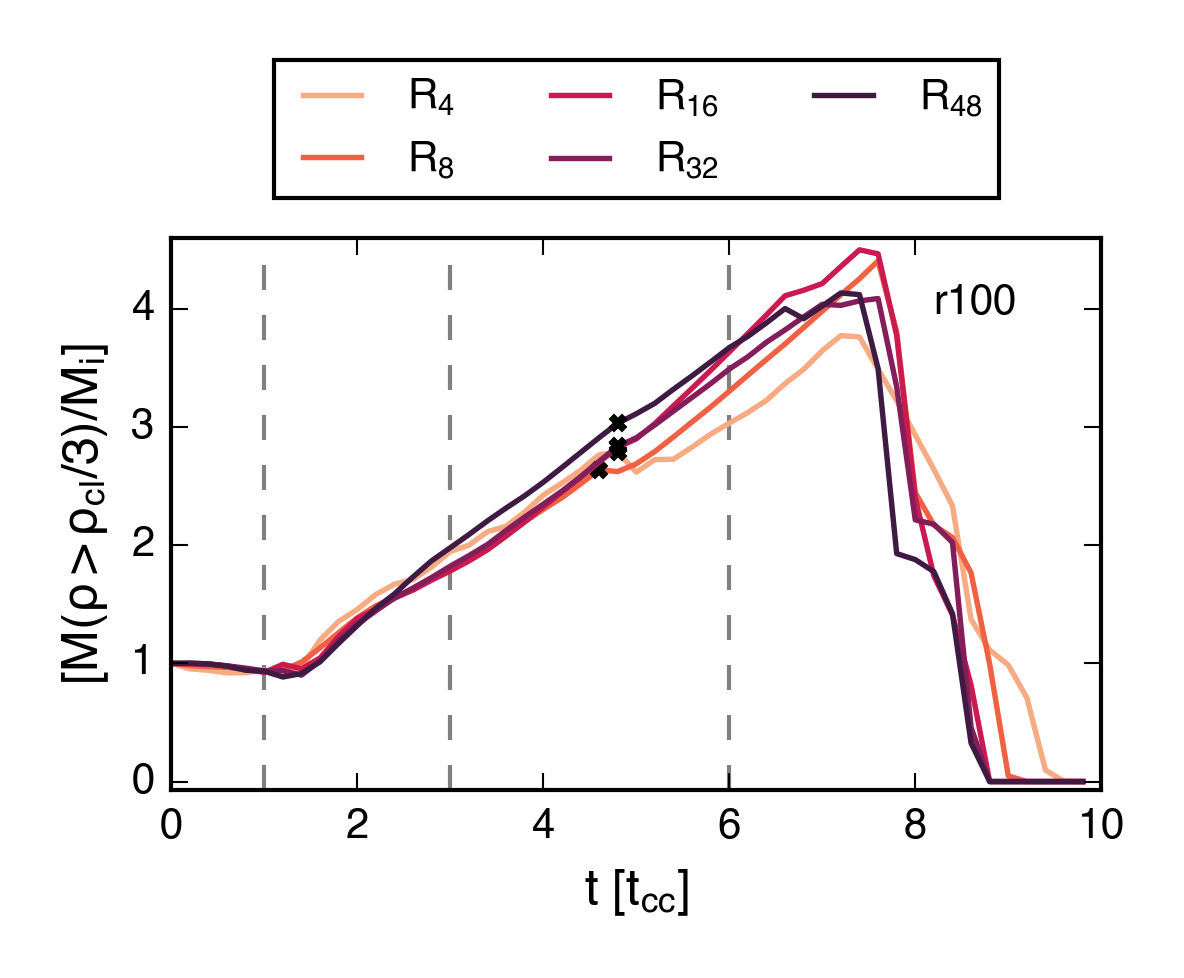}
    \includegraphics[width=\linewidth]{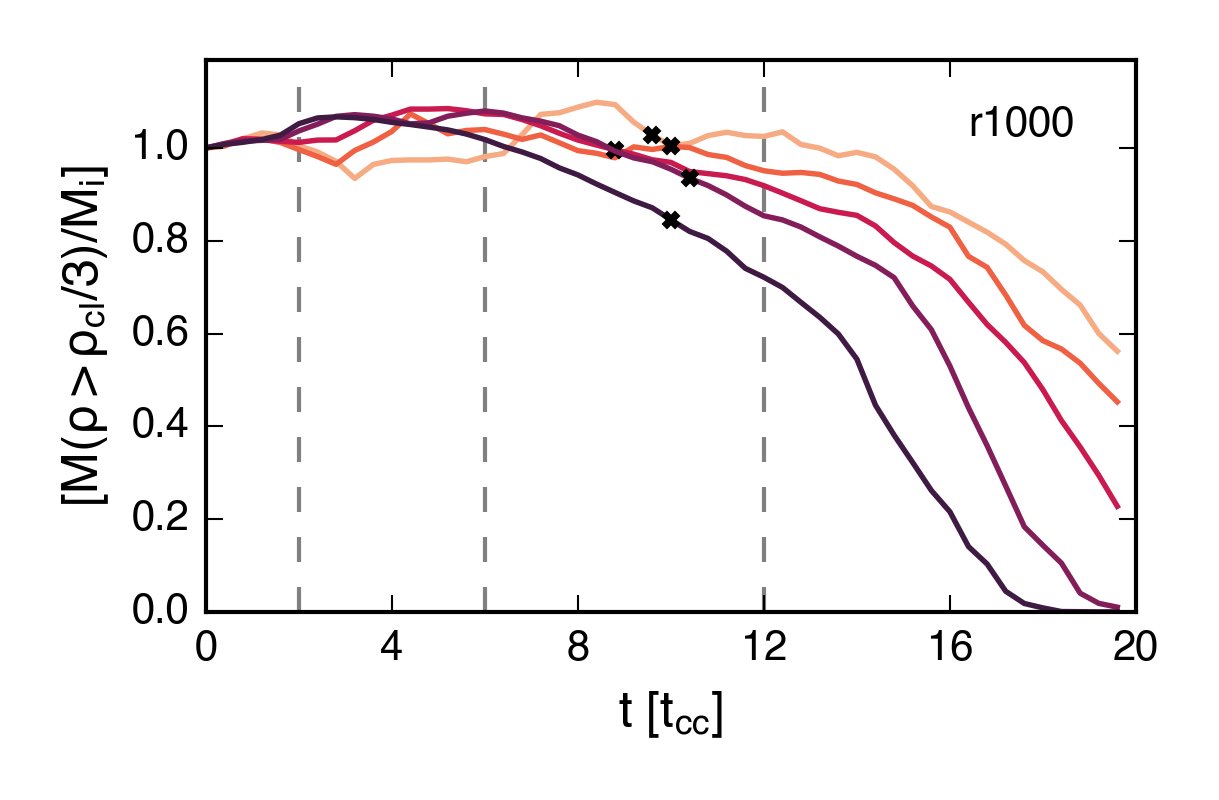}
    \caption{Change in cloud mass over time normalized to the initial cloud mass at resolutions $R_{\rm cl}/\Delta x$ = 4, 8, 16, 32, and 48. Top and bottom panels depict models r100 and r1000, respectively. Vertical lines correspond to snapshots shown in Figure \ref{fig:radiative snapshots}. Black X's mark the time when cloud mass begins to leave the box.}\label{fig:radiative mass}
\end{figure}

\begin{figure}[t!]
    \includegraphics[width=\linewidth]{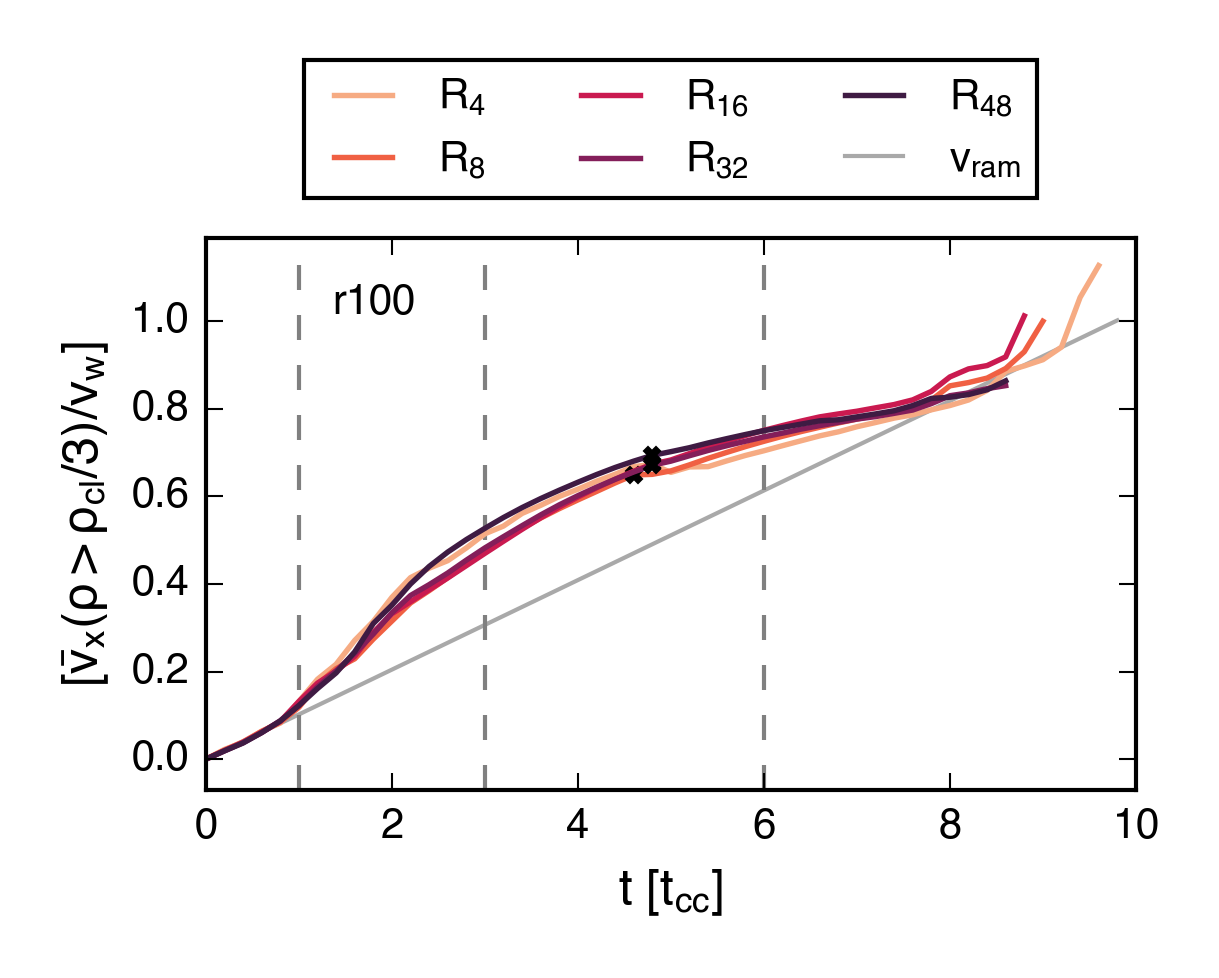}
    \includegraphics[width=\linewidth]{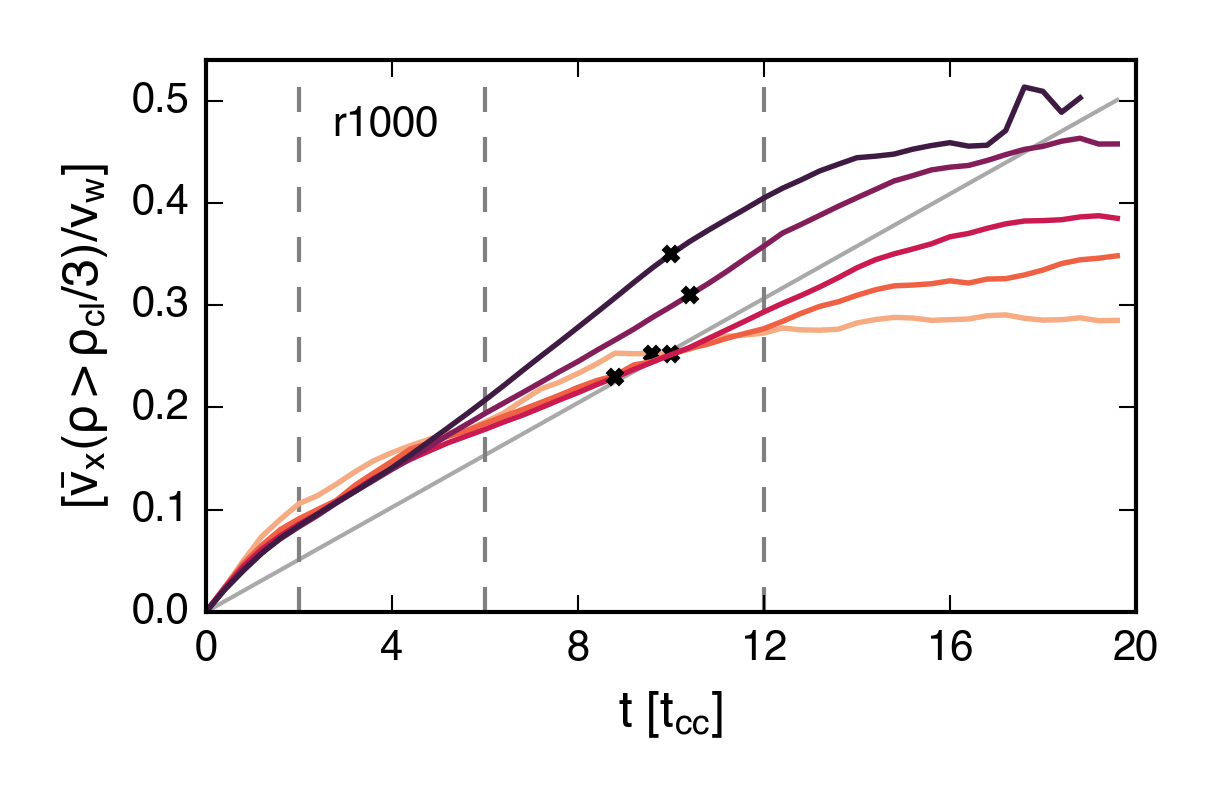}
    \caption{Change in cloud velocity over time normalized to the wind velocity at resolutions $R_{\rm cl}/\Delta x$ = 4, 8, 16, 32, and 48. Top and bottom panels depict models r100 and r1000, respectively. Vertical dashed lines correspond to snapshots shown in Figure \ref{fig:radiative snapshots}. Black X's mark the time when cloud mass begins to leave the box. The solid gray line in each panel indicates the expected cloud velocity due to ram pressure acceleration.} \label{fig:radiative velocity}
\end{figure}

Figure \ref{fig:radiative snapshots} shows column density snapshots for the radiative simulations. In the subsonic regime (top panel), the cloud evolution appears more similar across resolution, with the three highest resolution clouds advancing slightly farther in the box than the lowest two resolutions. The highest resolutions also appear to better capture small cloudlets along the tail of the cloud. 

Unlike in the adiabatic case, where hydrodynamic instabilities destroyed the clouds before they could reach the rightmost boundary of the simulation box, radiative cooling here facilitates cloud survival and growth, allowing cloud material above the density threshold to leave the box. Obviously, not capturing the material leaving the box will bias our results by removing the highest velocity cloud material from the overall average. Thus, our analysis in the following section will primarily be focused on times in the cloud evolution before the cloud material begins to exit the simulation boundary. Figure \ref{fig:radiative mass} shows the mass evolution for the radiative simulations. We indicate when cloud material in the tail begins to leave the box for each model with black X's; this occurs for all models at $4-5~t_{\rm cc}$. Before this, all models exhibit relatively consistent mass growth, reaching approximately 2.5 times the initial mass before material begins to exit the box. The clouds continue to grow as the bulk of the cloud is carried toward the edge of the box, reaching as much as 4 times the initial mass before most of the material exits, with the sharp decrease after $7~t_{\rm cc}$ showing when the bulk of the cloud moves out of the box. This dramatic growth is consistent with predictions of cloud survival, as the ratio of $t_{\rm cool,mix}/t_{\rm cc} \sim 10^{-3}$ puts the cloud safely in the survival/growth regime. Model $R_\mathrm{48}$ reaches slightly higher mass than the other models before cloud material starts leaving, indicating that mixing-induced growth may depend slightly on resolution, but overall, the growth trend is very similar across all resolutions, especially prior to cloud material leaving the box.

This lack of a resolution trend is further demonstrated in the top panel of Figure \ref{fig:radiative velocity}, which shows the average cloud velocities for the radiative simulations. Here we see that the velocity curves for all models track each other closely. At all resolutions, cloud material accelerates to over 60\% of the wind velocity before beginning to exit the box, and remaining material reaches $\sim80\%$ of the wind speed before exiting the simulation domain entirely. Given a sufficient box size, clouds in this regime would go on to reach entrainment in the wind at later times. As in the adiabatic case, the velocities of each resolution are indistinguishable from one another for the first $t_{\rm cc}$, and they follow the predicted velocity from the wind's ram pressure identically. After this point, the clouds in all models begin to accelerate more quickly, as the momentum contribution from mixing with the wind begins to take effect.

\subsection{Radiative Supersonic Models}

In the bottom panel of Figure \ref{fig:radiative snapshots}, we show snapshots of cloud evolution in the radiative supersonic regime at $2~t_{\rm cc}$, $6~t_{\rm cc}$, and $12~t_{\rm cc}$. We see a qualitatively different cloud structure relative to the subsonic case. In the subsonic models, the cloud material was largely smooth and contiguous, with only a few separate cloudlets in the tail. In the supersonic case, we see that the clouds are more broken up in all models, but particularly at the highest resolutions. This is consistent with the behavior seen in other works---in the supersonic wind, the increased wind speed means that the cloud is closer to the destruction regime ($t_{\rm cool,mix}/t_{\rm cc} \sim 10^{-2}$), where individual cloudlets in the tail are not always able to rejoin the bulk of the cloud. In the subsonic case our clouds were firmly in the growth regime. 

We also observe more differences with resolution in the supersonic radiative case. The bottom panel of Figure \ref{fig:radiative mass} shows cloud mass for the radiative supersonic models, and we see that for all resolutions, cloud mass stays relatively close to the initial mass before cloud material begins leaving the box around $9~t_{\rm cc}$. Initially, the highest resolution models appear to gain some mass while the lower resolution models lose some, but around $5~t_{\rm cc}$, this trend reverses, indicating that exact results for what constitutes cloud ``survival" in this regime may be quite sensitive to resolution and the numerical algorithm chosen. The mass in the higher resolution models steadily decreases after this, with the rate of mass loss increasing with increasing resolution. The ratio of $t_{\rm cool,mix}/t_{\rm cc}$ here predicts cloud survival, however as cloud mass begins to exit the box around 9 $t_{\rm cc}$, it is unclear whether the cloud would eventually exhibit mass growth.

In Figure \ref{fig:radiative velocity}, we see the same reversed resolution trend at $5~t_{\rm cc}$ for the velocity evolution. All resolutions initially have very similar velocities, with the low resolutions moving slightly faster than the higher resolutions, until the low resolutions begin to flatten; after $5~t_{\rm cc}$ velocity increases with increasing resolution. $R_{48}$ reaches the highest velocities before cloud material begins leaving the box, attaining 35\% of the wind velocity, while $R_{4}$ has only reached 25\% at this point. At later times, the higher resolution models consistently have higher velocity material, though we note that cloud material has left the box at this point. As in the a1000 model, we calculate the expected velocity from the ram pressure of the post-shocked wind. The actual early-time velocities exceed this prediction, likely because our simulations are under-resolving the post shock region, causing the cloud to experience a force that is a combination of the pre-shock and post-shock wind conditions.

\section{Discussion and Conclusions} \label{sec:disc}

\subsection{Cloud Acceleration Mechanisms}

In all simulations, the early-time evolution is similar between resolutions while the late-time behavior diverges. This shift suggests that the dominant acceleration mechanism of the cloud transitions from ram pressure-driven to mixing-driven, as mixing is sensitive to the smallest resolved scale, while ram pressure at early times before the cloud is disrupted only depends on the initial projected surface area and wind properties, which are independent of resolution. We present a picture where clouds are linearly accelerated by ram pressure for $ \sim1-2 ~ t_{\rm cc}$ until hydrodynamic instabilities are able to disrupt them. These instabilities drive additional acceleration via mixing, and can lead to mass growth in the radiative case. The early-time velocities of our clouds closely track the velocities predicted by a simple formula for ram pressure. While the surface area is the same for all of our cloud setups, the ram pressure scales as $v_{\rm w}^{2}$, so we expect the acceleration due to ram pressure to be greater in the supersonic models. In the adiabatic case, the cloud in the supersonic wind experiences a ram pressure force that is 25 times that of the subsonic wind, possibly explaining why the relative difference in velocities between resolutions is larger in a100 than in a1000. 

At later times, the adiabatic and radiative simulations show two different resolution trends, which may be a result of the difference in the morphology of the disrupted cloud in each scenario. In the radiative simulations, cooling of the mixed material in the cloud's wake leads to a low pressure region that keeps the cloud material largely contained in a narrow filamentary structure. In the adiabatic models, the disrupted cloud material can disperse orthogonally, and the cloud material in each resolution appears different morphologically (see Fig. \ref{fig:adiabatic snapshots} at $3~t_{\rm cc}$). For example, in the subsonic case, a plume is visible ahead of our $R_{\rm 16}$ cloud, likely an artifact of the Rayleigh-Taylor instability. We propose that $R_{\rm 16}$ experiences a combination of hydrodynamic instabilities that allows it to retain a cold dense core for a longer time. The non-monotonic resolution dependence of a100 thus appears to be a result of these morphological differences.  We anticipate that this effect would be less prominent with an initially non-homogeneous cloud, an investigation that we leave for future work.

\subsection{Resolving Cloud Evolution}

There has been considerable discussion about what resolution is sufficient for resolving the hydrodynamical interactions between the hot and cold phases. Radiative mixing layer studies have suggested that the interface layer between phases must be resolved with at least four cells \citep{Koyama2004, Robertson2010, Fielding2020}. Similarly, \cite{McCourt2018} argued that clouds larger than a characteristic ``shattering length" $l_{\rm shatter} = c_{\rm s}t_{\rm cool}$ will fragment into cloudlets of size $l_{\rm shatter}$, which in our simulations is of the sub-parsec scale. This could potentially enhance cloud mass growth by increasing the surface area available for mixing to occur, implying that for convergence in cloud mass growth, $l_{\rm shatter}$ must be resolved. However, in cloud-crushing simulations, it has been argued that convergence in bulk properties can be achieved by resolving the length scale of the cloud $\sim R_{\rm cl}$. For example, \cite{Gronke2019} demonstrated that mass growth was independent of resolution, regardless of whether $l_{\rm shatter}$ was resolved. They showed that mass growth was converged at their lowest resolution of 8 cells per $R_{\rm cl}$---a criterion that has been widely adopted in subsequent cloud-crushing studies. 

Our results demonstrate that 8 cells per $R_{\rm cl}$ is in fact more than sufficient to capture mass growth and acceleration in the low mach-number, fast-cooling regime. In fact, our simulations in this regime appear to be converged at only 4 cells per $R_{\rm cl}$, the lowest resolution we modeled (see top panels of Figures \ref{fig:radiative mass} and \ref{fig:radiative velocity}). However, while this resolution may be sufficient for capturing mass growth, it may be inadequate for resolving cloud destruction. Radiative mixing layer simulations show that entrainment of hot gas mass into the mixing layer is dominated by radiative cooling rather than the KH instability \citep{Ji2019}, whereas cloud destruction is driven by KH. Since the growth of KH is sensitive to the smallest resolved scale, the destruction regime should be more sensitive to resolution than the survival/growth regime. The lack of convergence in destruction times and velocity evolution in our adiabatic simulations supports this premise. In the radiative case, higher Mach number winds or slower cooling pushes clouds towards the destruction regime, where the 8-cell criterion may no longer hold. Indeed, the evolution of r1000 does not appear to be converged even at our highest resolution, 48 cells per $R_{\rm cl}$, and these clouds experience significantly more destruction than their counterparts in a subsonic wind. 

These findings have interesting implications for large-scale simulations, particularly cosmological simulations where individual cool clouds may be poorly resolved. Many recent studies have discussed the difficulty that cosmological simulations face in resolving the small-scale interactions between the cold and hot phases in the CGM and in galactic winds, and some have attempted to address this by explicitly improving resolution in the CGM \citep{Suresh2018, Hummels2019, Peeples2019, Rey2024}. Perhaps somewhat optimistically, our results indicate that in the subsonic, radiative cooling regime relevant in the low-redshift CGM, a resolution of only a few cells per cloud radius may be sufficient to adequately capture overall cloud acceleration and mass growth. If clouds in the CGM follow a mass function in which the bulk of the mass is in resolved clouds, this offers hope that some state-of-the-art cosmological simulations with CGM refinement may be accurately capturing the cold gas evolution. For example, in the GIBLE RF512 suite \citep{Ramesh2024}, a typical cell size in the CGM is approximately 100 pc, and clouds composed of at least 10 cells are considered in their analysis. A typical cloud in the CGM, where Mach numbers are low, will be safely in the growth regime. Thus, our results suggest that the overall properties of massive cold clouds in these simulations are robust to resolution. We note that the situation is somewhat more challenging for galaxy-scale simulations that attempt to resolve winds, like those in \cite{Hu2018, Smith2018, Schneider2024}, due to the higher pressures and prevalence of supersonic outflows. At higher Mach numbers, cold gas evolution appears to be more sensitive to resolution, and higher resolutions will be necessary for convergence.

\subsection{Additional Physics}

In this paper, we have considered resolution effects in idealized cloud-wind simulations that include radiative cooling only. The inclusion of additional physics--such as thermal conduction or magnetic fields--is known to modify the evolution of multiphase gas and may alter the resolution effects we see. For example, \cite{Koyama2004} showed that in one-dimensional simulations of thermal instability, the structure and turbulent velocity of the multiphase medium is strongly dependent on the numerical resolution. However, the inclusion of thermal conduction reduced these resolution effects when the characteristic length scale of the thermal conduction, the Field length, was resolved by at least 3 cells. Additionally, the inclusion of magnetic fields has been shown to inhibit cloud fragmentation \citep{McCourt2015}, a process that becomes more pronounced at higher resolutions. We therefore might expect the inclusion of magnetic fields to lessen the resolution effects seen in purely radiative simulations. This would be most important for clouds in or near the destruction regime ($t_{\rm cool, mix} \geq t_{\rm cc}$) that experience more fragmentation, for example our r1000 simulations, where significant fragmentation is visible later in the cloud's evolution and increases at higher resolutions (Fig. \ref{fig:radiative snapshots}, bottom panel).

\subsection{Summary}

Our simulations show that the effect of numerical resolution on the evolution of cloud-crushing simulations depends on whether the cloud exists in the growth or destruction regime. We find that cloud mass growth and acceleration are converged at a resolution of 4 cells per cloud radius in simulations with fast cooling and a subsonic wind. This implies that the current highest resolution cosmological simulations may be capable of accurately capturing the bulk of cold gas evolution in the CGM. The lack of convergence in all of our other models, however, implies that this criterion is not universal, and does not appear to hold in the destruction regime. Our radiative cloud in a supersonic wind displays a strong monotonic relationship with resolution, where clouds are destroyed and accelerated increasingly quickly as the resolution increases, demonstrating that the relationship to resolution is sensitive to the Mach number. In the adiabatic regime, the dependence on resolution is unexpectedly non-monotonic. What constitutes a sufficient resolution for convergence in properties like cloud acceleration and survival, then, will vary greatly for each cloud-crushing setup, and caution should be taken when adopting a resolution choice.

\begin{acknowledgments}
We wish to thank the referee for their helpful comments and suggestions which have improved this work. H.J.L thanks Matthew Abruzzo, Robert Caddy, Alwin Mao, and Orlando Warren for many helpful discussions. This research was supported in part by the University of Pittsburgh Center for Research Computing, RRID:SCR\_022735, through the resources provided. Specifically, this work used the H2P cluster, which is supported by NSF award number OAC-2117681. This research used resources of the Oak Ridge Leadership Computing Facility, which is a DOE Office of Science User Facility supported under Contract DE-AC05-00OR22725, using Frontier allocation AST181. E.E.S. acknowledges support from NASA ATP grant 80NSSC22K0720 and the David and Lucile Packard Foundation (grant no. 2022-74680).

\end{acknowledgments}

\vspace{5mm}

\software{Cholla \citep{Schneider2015}, \texttt{numpy} \citep{VanDerWalt2011}, \texttt{matplotlib} \citep{Hunter07},  \texttt{hdf5} \citep{hdf5} \texttt{seaborn} \citep{Waskom2021}}

\bibliography{res-dependence}{}

\appendix 

\section{Density Cut} \label{sec:cuts}

\begin{figure*}
\centering
\includegraphics[width=0.45\linewidth]{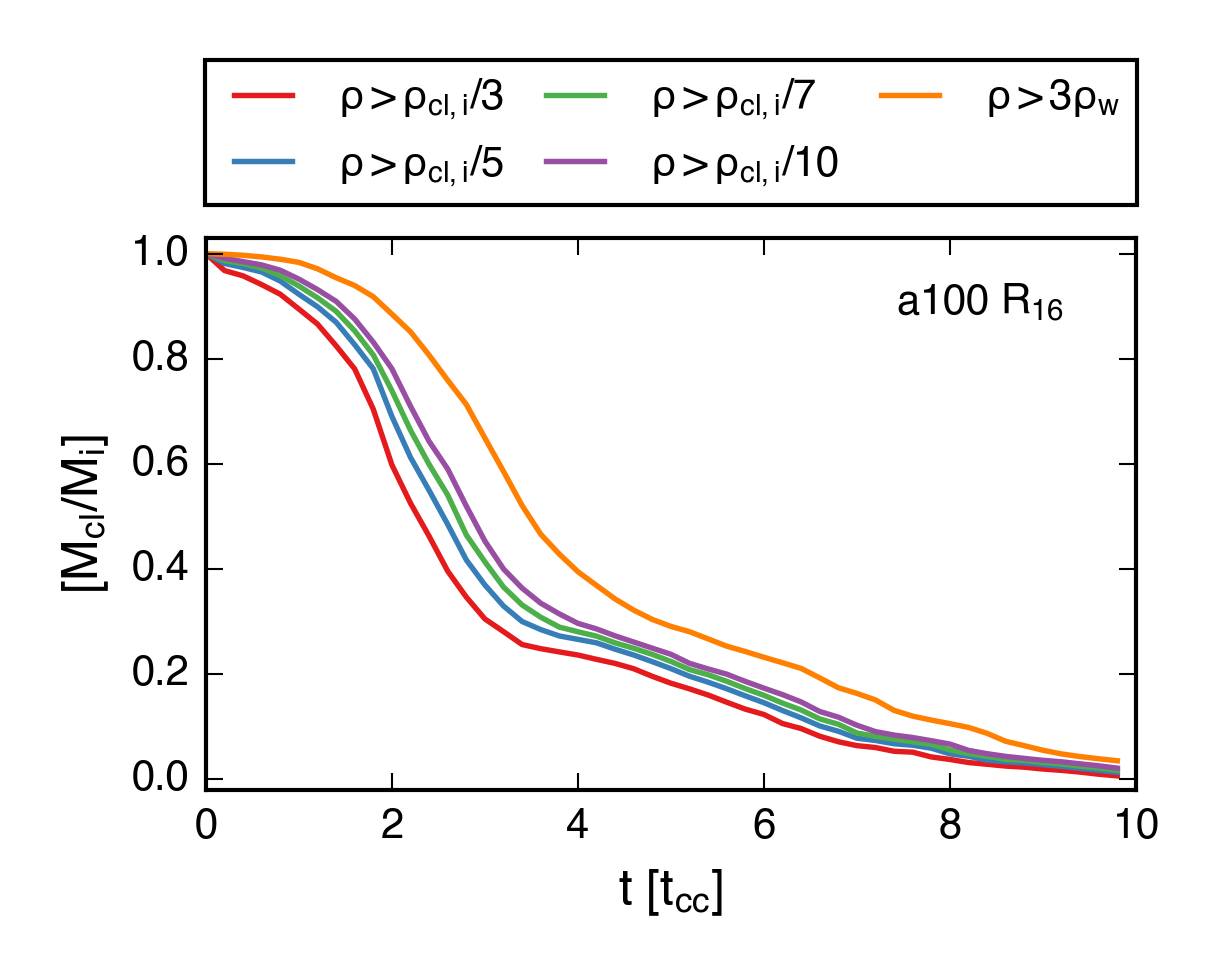}
\includegraphics[width=0.45\linewidth]{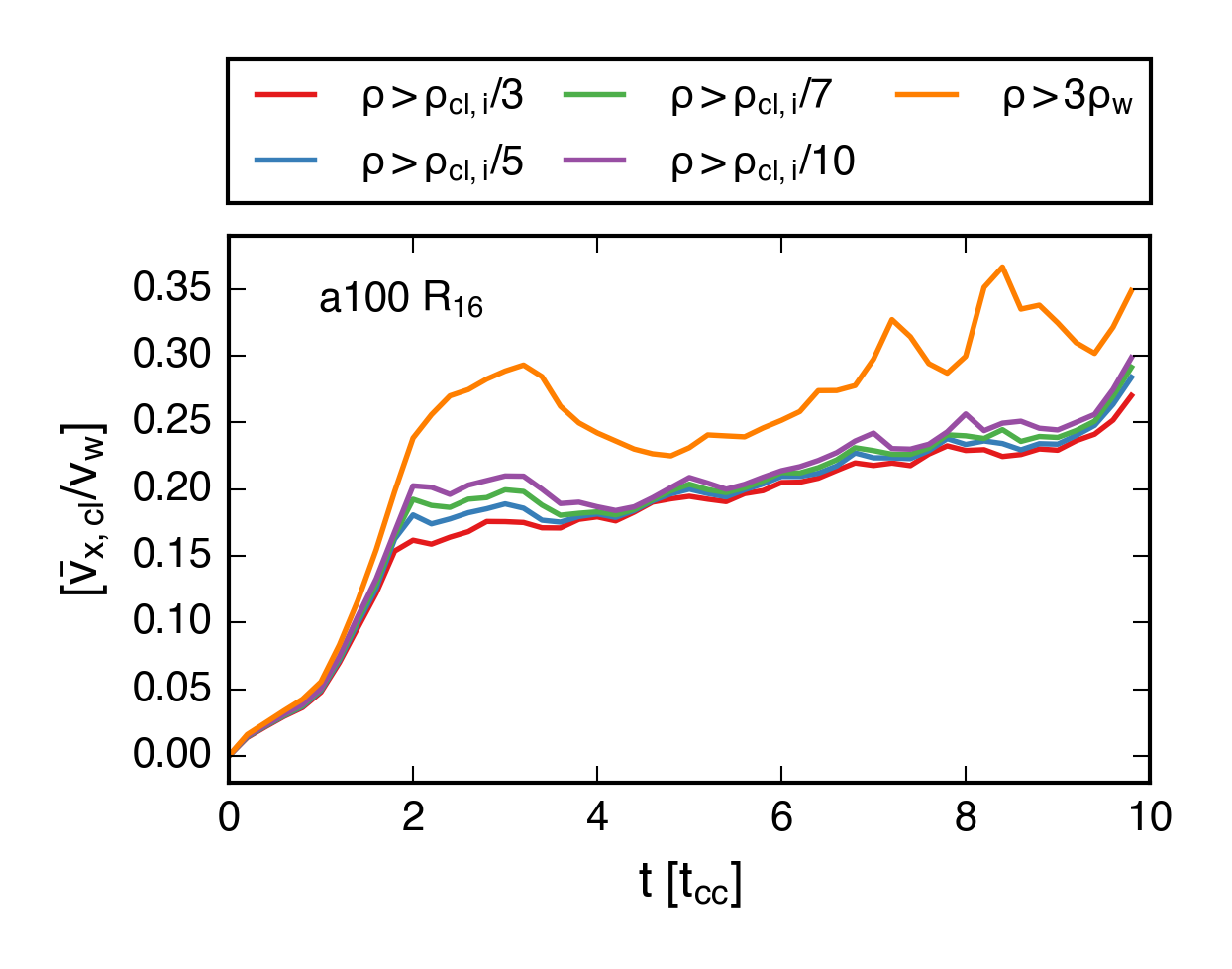} 
\caption{Results for various density cuts used to define        ``cloud material", for our a100 model at a resolution of $R_{\rm 16}$. Left panel: change in cloud mass over time, normalized by the initial cloud mass. Right panel: Change in average cloud velocity over time, normalized by the wind speed.}\label{fig:density cuts}
\end{figure*}

In this appendix we demonstrate the effects of using different density criteria to define what cells constitute cloud material in our analysis. In the left panel of Fig. \ref{fig:density cuts} we plot the mass evolution of our a100 cloud at a resolution of 16 cells per cloud radius, where we define cloud material with five different density cuts: cells with a density greater than 1/3, 1/5, 1/7, and 1/10 of the initial cloud density, as well as cells with densities greater than just above that of the wind ($3 \times \rho_{\rm w}$). As the density criteria gets smaller, the total mass increases slightly as expected, with our choice of 1/3 yielding $\sim$ 70\% of the initial mass by 2 $t_{\rm cc}$, and a density above 3 times that of the wind yielding $\sim$ 90\% at the same time. However, the trends are qualitatively very similar and the cloud is destroyed at a similar time in all cases. 

In the right panel, we plot the velocity evolution for each density cut. Consistent with the mass evolution, we calculate slightly higher velocities with smaller density fractions, but the highest four density cuts generally converge after 4 $t_{\rm cc}$ We see the greatest discrepancy when accounting for more of the mixed gas: at 3 $t_{\rm cc}$ our cloud with a density cut of $\rho_{\rm cl,i}/10$ is moving with a velocity of 20\% of the wind speed, and our cloud with a density cut of $3\rho_{\rm w}$ is moving at 30\% of the wind speed. This confirms that our density cut of 1/3 is somewhat biased against mixed gas with lower densities that may have higher velocities.

\end{document}